# LITHIUM TANTALATE BULK ACOUSTIC RESONATOR FOR PIEZOELECTRIC POWER CONVERSION

Ziqian Yao[1], Clarissa Daniel[2], Eric Stolt[2], Vakhtang Chulukhadze[1], Juan Rivas Davila[2], and Ruochen Lu[1]
[1]Department of Electrical and Computer Engineering, The University of Texas at Austin, Austin, USA
[2]Department of Electrical Engineering, Stanford University, Stanford, USA

## ABSTRACT

We present the first lithium tantalate (LT) thickness-extensional (TE) mode bulk acoustic resonators for piezoelectric power conversion, showcasing a low temperature coefficient of frequency (TCF) of −13.56 ppm/K. These resonators also exhibit high quality factors ($Q$) of 1698, and electromechanical coupling coefficients ($k^2$) of 8.8%, making them suitable for efficient power conversion applications. A grounded ring structure is leveraged for spurious mode-free response and figure-of-merit (FoM=$k^2 \cdot Q$) enhancement near the series resonance, within the power converter's operational range. The temperature dependency of $Q$ and $k^2$ are experimentally tested over a wide temperature range, from 25 °C to 130 °C, demonstrating the resonator's thermal stability and consistent performance under varying conditions. This work highlights the potential of LT resonators in the future development of thermally stable power electronic systems, enabling more reliable and efficient piezoelectric power converters.

## KEYWORDS

Acoustic resonator, lithium tantalate, piezoelectric power conversion, temperature coefficient of frequency.

## INTRODUCTION

Piezoelectric power conversion has gained attention as an emerging approach for replacing magnetic inductors with acoustic resonators in power converters [1]. Conventional magnetic inductors are challenging to miniaturize due to the inherently large wavelength of electromagnetic (EM) waves and significant parasitic effects at higher switching frequencies [1], [2], [3], [4]. In contrast, acoustic resonators convert electrical energy into stored mechanical vibrations, offering significant advantages in both size reduction and efficiency improvement [5], drawing on principles similar to those used in miniature acoustic resonators for commercially successful radio frequency (RF) filters[5] .

The state-of-the-art (SoA) piezoelectric materials for power conversion (Table 1) have been evaluated by analyzing key performance metrics, including electromechanical coupling coefficient ($k^2$), quality factor ($Q$), and thermal stability. PZT-based resonators demonstrate strong coupling efficiency, with $k^2$ reaching 30-40%, but suffer from relatively high loss and poor thermal stability[6], [7]. Lithium niobate (LN) thickness extensional (TE) mode resonators, where vibration occurs primarily along the thickness direction of the piezoelectric material, have demonstrated advantages with their high FoM [8] and great linearity, enabling compact power converters up to 3.2 kW [9]. However, their large temperature coefficient of frequency (TCF) around −70 ppm/K rendered them unsuitable for applications requiring high thermal stability.

| Platform | Ref. | $f_s$ (MHz) | $k_t^2$ | Q | TCF (ppm/K) |
|---|---|---|---|---|---|
| PZT-TE [6] | Boles et al | 0.61 | 31% | 2500 | N/A |
| PZT-Radial [7] | Daniel et al | 0.174 | 42.5% | 1198 | -35 |
| LN-TE [11] | Braun et al | 6.82 | 29% | 4178 | N/A |
| LN-TE [12] | Touhami et al | 6.28 | 25.5% | 3700 | N/A |
| LN-TS [2] | Nguyen et al | 5.94 | 45% | 3500 | N/A |
| LN-TE [7] | Daniel et al | 10.1 | 29.5% | 3500 | -69 |
| LT-TE (This work) | Yao et al | 6.5 | 8.8% | 1968 | -13.56 |

*Table 1. Comparison of this work to prior art. This work features low TCF and good FoM as resonators, promising for power conversion applications.*

| Material | Mode | e | $\varepsilon_r$ | FoM [$e^2/(\varepsilon \cdot c)$] | Loss Tangent | Curie Temp. (°C) | TCF (ppm/K) | Wafer Available |
|---|---|---|---|---|---|---|---|---|
| PZT-5A | LE | 5.35 | 826 | 0.33% | 0.018 | 320°C | −35 | Y |
| AlN | LE | 0.59 | 10.3 | 0.96% | 0.001 | 1150°C | −28 | Y |
| AlN | TE | 1.47 | 10.3 | 6.0% | 0.001 | 1150°C | −28 | Y |
| Sc$_{0.3}$Al$_{0.7}$N | LE | 0.70 | 21.1 | 0.84% | 0.0036 | 1150°C | −35 | N |
| Sc$_{0.3}$Al$_{0.7}$N | TE | 2.38 | 21.1 | 9.7% | 0.0036 | 1150°C | −35 | N |
| LN | LE | 1.83 | 38.6 | 4.8% | 0.00013 | 1200°C | -70 | Y |
| LN | TE | 4.53 | 38.6 | 31.2% | 0.00013 | 1200°C | -69 | Y |
| LT | TE | 3.25 | 41.7 | 10.7% | 0.00013 | 600°C | -12 | Y |

*Table 2. Comparison between LT and other acoustic platforms for piezoelectric power converters. LT features a combination of stable temperature coefficient (low TCF), low EM loss, high coupling, and commercially available wafers.*

In this work, we address the limitations of existing piezoelectric platforms by utilizing lithium tantalate (LT), which offers a balanced combination of high figure of merit (FoM), excellent piezoelectric properties, low electromagnetic loss, and improved thermal stability (Table 2). More specifically, commonly utilized piezoelectric materials, such as LN and PZT, possess poor thermal stability, with TCF values reaching −69 ppm/K (LN-TE) and −35 ppm/K (PZT-5A). Comparatively, LT exhibits a much-improved TCF of −12 ppm/K, ensuring greater frequency stability across varying temperature ranges. In addition to the low TCF, 36Y LT TE mode resonators achieve high FoM of 10.7%, surpassing materials such as aluminum nitride (AlN) and scandium-doped aluminum nitride ($Sc_{0.3}Al_{0.7}N$). Furthermore, unlike AlN/ScAlN, which lacks commercially available wafer options for MHz power converters, LT is readily available, making it more viable for scalable device fabrication. By leveraging these advantages, LT emerges as a superior alternative to existing piezoelectric materials, particularly in applications requiring stable frequency operation, low energy loss, and enhanced power conversion efficiency [10] .

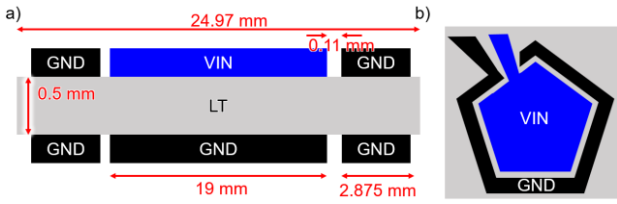

*Fig. 1 Schematics of LT thickness extensional (TE) mode bulk acoustic resonators with spurious suppression grounded ring. The key dimensions and electrical configurations are labeled in the figure.*

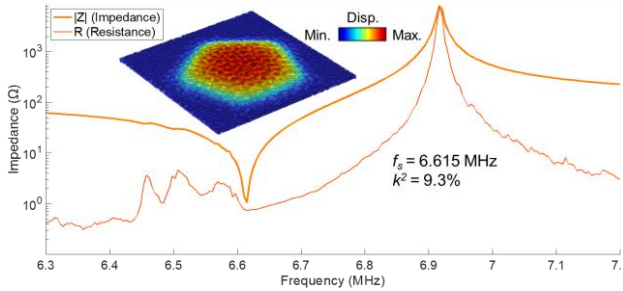

*Fig. 2. Finite element simulated frequency domain impedance and resistance, along with inset figure showing the displacement mode shape at resonance*

**DESCRIPTION OF THE NEW METHOD**

In this work, we implemented a spurious mode free resonator with a grounded ring design to effectively suppress spurious modes and enhance performance near the series resonance frequency ($f_s$) [9]. The schematic of LT resonator design is shown in Fig. 1. The device consists of a pair of pentagon-shaped electrodes placed on both sides of an LT wafer with a film thickness of 0.5 mm. The wafer thickness was chosen based on the frequency specifications set by the desired power converter operation (around 6.5 MHz). The resonator is 24.97 mm in length, with the top electrode featuring a central VIN region, which is surrounded by a grounded ring that is 2.875 mm wide. A 0.11 mm gap separates the VIN region from the grounded ring. On the bottom side, the electrode design mirrors the top structure, with three distinct grounded regions aligned to match the shape of the top pentagonal shape electrode. The pentagon shape was chosen as it provides better spurious mode smoothing compared to a circular design, improving overall resonator performance. Notably, this is the first pentagonal resonator designed for piezoelectric power conversion applications, demonstrating its potential in high-efficiency energy transfer systems.

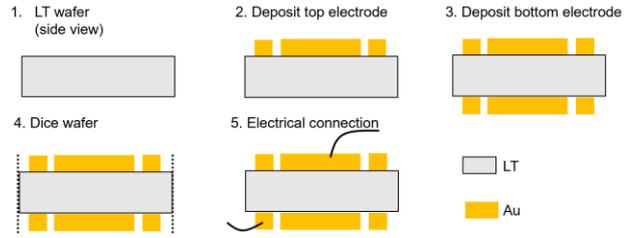

*Fig. 3 Fabrication and integration process flow.*

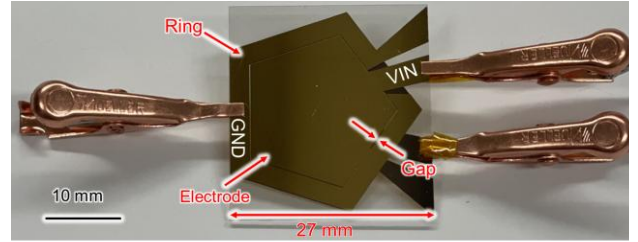

*Fig. 4 Fabricated resonators for power electronic testing at room and elevated temperatures.*

To validate our design, we performed finite element analysis (FEA) using COMSOL to simulate the admittance and displacement mode shapes (Fig. 2). From the simulated mode shape, we observed that the pentagon ring structure vibrates uniformly, with the highest displacement amplitude concentrated near the center. The results indicate that our proposed design, which incorporates the grounded pentagon ring structure, exhibits an impedance and resistance response that is nearly spurious-free. This design can effectively extend the operational range of the converter. Based on our analysis, we extracted a series resonance frequency ($f_s$) of 6.615 MHz and an electromechanical coupling coefficient ($k^2$) of 9.3%, demonstrating the effectiveness of the grounded ring structure in improving resonator performance.

**DEVICE FABRICATION**

After our design is thoroughly validated in COMSOL, the TE resonator is fabricated and integrated with a standard cleanroom process flow using a 3-in 0.5 mm thick 36Y LT wafer (Fig. 3). The wafer is first prepared and cleaned using acetone and isopropyl alcohol (IPA). After spin-coating the photoresist, the top electrodes are defined through a standard photolithography process, followed by electron beam metal evaporation. Gold (Au) is selected for its excellent electrical conductivity. To ensure proper adhesion, a thin 10 nm titanium (Ti) layer is deposited before the 300 nm Au layer. Following the metal lift-off process, the wafer is flipped to the backside, and the cleaning and metal deposition steps are repeated, prior to backside alignment. This ensures that the ring structure is properly grounded while preventing unintended electrical feedthrough between different structures. Finally, the individual fabricated resonators are diced from the wafers by using an automated wafer dicing saw.

The fabricated device is shown in Fig. 4, after being connected to clips for a performance test with a network analyzer and a thermally controlled chamber for testing at room and elevated temperatures. This setup enables precise characterization of the device's electrical and thermal behavior under varying conditions. The fabricated device is shown in Fig. 4, after being connected to clips for performance test with a network analyzer and a thermally controlled chamber for testing at room and elevated temperatures. This setup enables precise characterization of the device's electrical and thermal behavior under varying conditions.

**EXPERIMENTAL RESULTS**

The measured admittance response, presented in Fig. 5, reveals $f_s$ at 6.5 MHz, high $Q$ of 1700 and $k^2$ of 8.8%. The low resistance of approximately 800 mΩ and the spurious-free spectrum above $f_s$ further validate the effectiveness of the grounded ring design in spurious mode suppression and performance enhancement. These results are comparable in prior studies using PZT and LN (Table 2), demonstrating the competitiveness of the LT approach while possessing a low TCF.

To assess thermal stability, the device is then tested at elevated temperatures, gradually increasing from 25 °C to 130 °C, as illustrated in Fig. 6 (a). The extracted resonance frequencies across this temperature range are summarized in Fig. 6 (b), highlighting a low TCF of −13.56 ppm/K, much lower than that in PZT (−35 ppm/K) and LN (−70 ppm/K). To mitigate higher TCF and prevent runaway failure, open-loop control can be implemented [13]. However, this approach increases system complexity and poses the risk of accidentally driving the device into lateral spurious modes near series resonance ($f_s$), which could break the device.

Additionally, the variations of $k^2$ and $Q$ at different temperatures are also analyzed and plotted in Fig. 7 (a)(b), respectively. A slight performance degradation was observed, particularly in the $k^2$ values, which was attributed to the higher TCF of the shunt resonance ($f_p$). The reduction in $Q$ can be

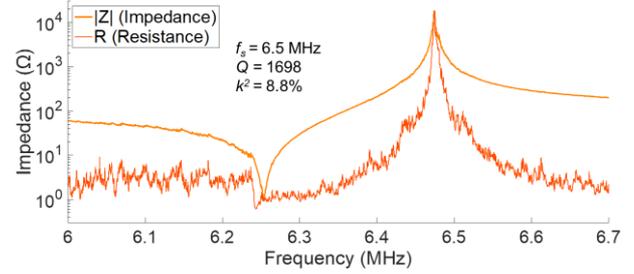

*Fig. 5 (a) Measured frequency domain impedance and resistance of fabricated LT resonator, featuring spurious-free spectrum above the resonance at 6.5 MHz, high Q of 1698 and $k^2$ of 8.8%.*

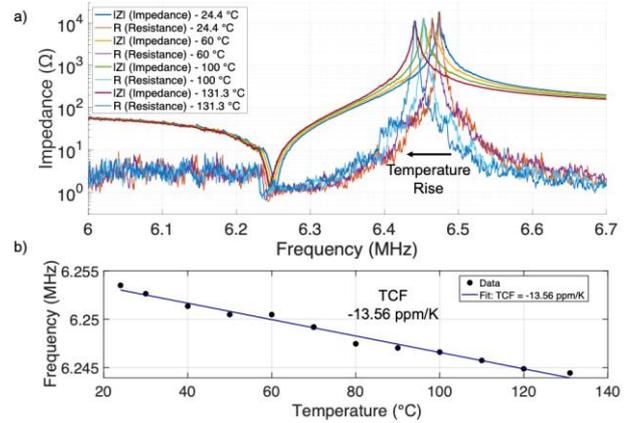

*Fig. 6 (a) Measured frequency domain impedance and resistance at different temperatures, along with (b) extracted TCF between 25 °C and 130 °C, showing low TCF of −13.56 ppm/K, much lower than that in PZT (−35 ppm/K) and LN (−70 ppm/K).*

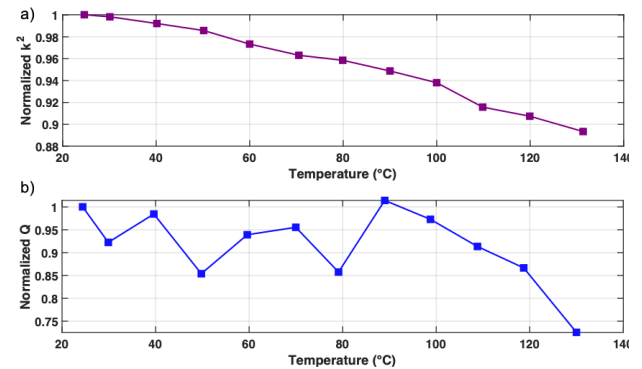

*Fig. 7 (a) $k^2$ and (b) Q at different temperature, extracted from the frequency domain measurement.*

partially caused by the increased series resistance ($R_s$) of the electrode material (Au), which increases at higher temperature due to the temperature-dependent resistivity of gold. Despite minor degradations, the resonator maintains excellent stability near the series resonance ($f_s$), where the piezoelectric power converters operate. This thermal resilience makes the LT resonator a promising candidate for high-power, thermally stable piezoelectric power conversion systems. In comparison with prior works, as detailed in Table 2, this study offers a significant advancement in achieving both high performance and thermal stability in piezoelectric resonators.

## CONCLUSION

In this work, we successfully demonstrated the LT TE mode bulk acoustic resonators for piezoelectric power conversion. The resonator achieves a low TCF of −13.56 ppm/K, a high $Q$ of 1698, and $k^2$ of 8.8%, making it well-suited for high-power and thermally stable power converter applications. The grounded ring structure effectively suppresses spurious modes and enhances the FoM, ensuring stable operation near the series resonance frequency. Thermal testing from 25 °C to 130 °C confirmed strong thermal stability and consistent performance, outperforming conventional PZT and LN resonators in key aspects. These results demonstrate the potential of LT resonators as a competitive and reliable solution for efficient piezoelectric power conversion systems.

## ACKNOWLEGEMENTS

The authors would like to thank DARPA HOTS project for funding support.

## CONTACT

*Ziqian Yao, hanson.yao@utexas.edu